\documentclass[a4paper,11pt]{amsart}
\usepackage{amsmath, amsthm, amssymb, amscd, mathrsfs, eucal,
  amsfonts, latexsym}

\newcommand{\calA}{{\mathcal A}}

\newcommand{\calD}{{\mathcal D}}

\newcommand{\calH}{{\mathcal H}}
\newcommand{\calK}{{\mathcal K}}

\newcommand{\euA}{{\mathscr A}}
\newcommand{\euD}{{\mathscr D}}
\newcommand{\Obs}{\textbf{\textsf{Obs}}}
\newcommand{\Loc}{\textbf{\textsf{Loc}}}

\newcommand{\supp}{\mbox{supp}}

\newcommand{\obj}{\mbox{\rm obj}}
\newcommand{\Hom}{\mbox{\rm hom}}

\newcommand{\bC}{{\mathbb C}}
\newcommand{\gb}{\boldsymbol{g}}

\theoremstyle{plain}
\newtheorem{Thm}{Theorem}[section]

\newtheorem{Def}[Thm]{Definition}
\theoremstyle{remark}

\begin{document}

\author[Algebraic approach to quantum field theory]{Romeo Brunetti and Klaus 
Fredenhagen\\
II Inst. f. Theor. Physik, Univ. Hamburg\\
149 Luruper Chaussee, D-22761 Hamburg, Germany}

\title{Algebraic Approach to Quantum Field Theory}

\maketitle

\section{Introduction}

Quantum Field Theory may be understood as the incorporation of the principle of locality, which is at 
the basis of classical field theory, into quantum physics. There are, however, severe obstacles against a 
straightforward translation of concepts of classical field theory into quantum theory, 
among them the notorious divergences of 
quantum field theory and the intrinsic non-locality of quantum physics. Therefore, the concept of locality is 
somewhat obscured in the formalism of quantum field theory as it is typically exposed in textbooks. 
Nonlocal concepts as the vacuum, the notion of particles or the S-matrix play a fundamental r\^ole, and 
neither the relation to classical field theory nor the influence of background fields can be
properly treated.

Algebraic quantum field theory (synonymously, Local Quantum Physics), on the contrary, aims at emphasizing the concept of locality at every instance.
As the nonlocal features of quantum physics occur at the level of states (``entanglement''), not at the level of
observables, it is better not to base the theory on the Hilbert space of states but on the
algebra of observables. Subsystems of a given system then simply correspond to subalgebras of a given algebra.
The locality concept is abstractly encoded in a notion of independence of subsystems; two subsystems are 
independent if the algebra of observables which they generate is isomorphic to the tensor product
of the algebras of the subsystems.

Spacetime can then -- in the spirit of Leibniz -- be considered as an ordering device for systems.
So one associates to regions of spacetime the algebras of observables which can be measured in the pertinent
region, with the condition that the algebras of subregions of a given region can be identified
with subalgebras of the algebra of the region.

Problems arise if one aims at a generally covariant approach in the spirit of General
Relativity. Then, in order to avoid pitfalls like in the "hole problem", systems corresponding
to isometric regions must be isomorphic. Since isomorphic regions may be embedded into
different spacetimes, this amounts to a simultaneous treatment of all spacetimes of a suitable class.
We shall see that category theory furnishes such a description, where the objects are the systems 
and the morphisms the embeddings of a system as a subsystem of other systems.

States arise as secondary objects via Hilbert space representations, or directly as linear functionals 
on the algebras of observables which can be interpreted as expectation values and are therefore
positive and normalized. It is crucial that inequivalent representations (``sectors'') can occur, and 
the analysis of the structure of the sectors is one of the big successes of Algebraic Quantum Field Theory.
One also can study the particle interpretation of certain states as well as (equilibrium and non-equilibrium) 
thermodynamical properties. 

The mathematical methods in Algebraic Quantum Field Theory are mainly taken from the theory of operator
algebras, a field of mathematics which developped in close contact to mathematical physics, in particular 
to Algebraic Quantum Field Theory. Unfortunately, the most important field theories, from the point of
view of elementary particle physics, as Quantum Electro Dynamics or the 
Standard Model could not yet be constructed 
beyond formal perturbation theory with the annoying consequence that it seemed that the concepts of 
Algebraic Quantum Field Theory could not be applied to them. But recently, it was shown that formal perturbation
theory can be reshaped in the spirit of Algebraic Quantum Field Theory such that the algebras of observables 
of these models can be constructed as algebras of formal power series of Hilbert space operators. 
The price to pay is that the deep mathematics of operator algebras cannot be applied, but the crucial features
of the algebraic approach can be used.

Algebraic Quantum Field Theory was originally proposed by Haag as a concept by which scattering of particles
can be understood as a consequence of the principle of locality. It was then put into a mathematically
precise form by Araki, Haag and Kastler. After the analysis of particle scattering by 
Haag and Ruelle and the
clarification of the relation to the Lehmann-Symanzik-Zimmermann formalism by Hepp, the structure of
superselection sectors was studied first by Borchers and then in a fundamental series of papers by 
Doplicher, Haag and Roberts, soon after Buchholz and Fredenhagen established the relation to particles, and finally
Doplicher and Roberts uncovered the structure of superselection sectors as the dual of a compact group
thereby generalizing the Tannaka-Krein Theorem of characterization of group duals.

With the advent of two-dimensional Conformal Field Theory new models were constructed and it was shown 
that the DHR analysis can be generalized to these models. 
Directly related to conformal theories is the algebraic approach to
holo\-graphy in anti-de Sitter (AdS) spacetime by Rehren.

The general framework of Algebraic Quantum Field Theory may be described as a covariant functor between 
two categories. The first one contains the information on local relations and is crucial for the 
interpretation. Its objects are topological spaces with additional structures (typically globally 
hyperbolic Lorentzian spaces, possibly spin bundles with connections, etc.), its morphisms structure 
preserving embeddings. In the case of globally hyperbolic Lorentzian spacetimes one requires that the 
embeddings are isometric and preserve the causal structure. The second category describes the algebraic 
structure of observables, in quantum physics the standard assumption is that one deals with the category
of C$^*$-algebras where the morphisms are unital embeddings. In classical physics one looks instead at Poisson
algebras, and in perturbative quantum field theory one admits algebras which possess nontrivial representations
as formal power series of Hilbert space operators. It is the leading principle of Algebraic Quantum 
Field Theory that the functor $\euA$ contains all physical information. In particular, two theories are 
equivalent if the corresponding functors are naturally equivalent.

In the analysis of the functor $\euA$ a crucial r\^{o}le is played by natural transformations from other functors
on the locality category. For instance, a field $A$ may be defined as a natural transformation from 
the category of test function spaces to the category of observable algebras via their functors related to the
locality category.

\section{Quantum Field Theories as Covariant Functors}

The rigorous implementation of the generally covariant locality principle uses the language 
of category theory.

The following two categories are used :  
\begin{description}
\item[$\Loc$] The class of objects
$\obj(\Loc)$ is formed by all (smooth) $d$-dimensional 
($d\ge 2$ is held fixed), globally
hyperbolic Lorentzian spacetimes $M$ which are oriented and time-oriented.
Given any two such objects $M_1$ and $M_2$, the morphisms
$\psi\in\Hom_{\Loc}(M_1,M_2)$ 
are taken to be the isometric embeddings
$\psi: M_1 \to M_2$ of $M_1$ into
$M_2$ but with the following constraints;  
\begin{itemize}
\item[$(i)$] if $\gamma : [a,b]\to M_2$ is any 
causal curve and
$\gamma(a),\gamma(b)\in\psi(M_1)$ then the whole curve must be 
in the image $\psi(M_1)$, i.e., $\gamma(t)\in\psi(M_1)$ for all $t\in ]a,b[$; 
\item[$(ii)$] any morphism preserves orientation and
  time-orientation of the embedded spacetime.
\end{itemize} 
Composition is composition of maps, the unit element in $\Hom_{\Loc}(M,M)$ is given by 
the identical embedding ${\rm id}_M : M\mapsto M$ for
any $M\in\obj(\Loc)$.
\end{description}

\begin{description}
\item[$\Obs$] The class of objects
  $\obj(\Obs)$ is formed by all C$^*$-algebras possessing unit
  elements, and the morphisms are faithful (injective) unit-preserving
  $*$-homomorphisms. 
  The composition
  is again defined as the composition of maps,
  the unit element in
  $\Hom_{\Obs}(\calA,\calA)$ is for any $\calA \in \obj(\Obs)$ given
  by the identical map ${\rm id}_{\calA}: A \mapsto A$, $A\in\calA$.
\end{description}

The choice of the categories is done for definitiveness. One may
envisage changes according to particular needs, as for instance in
perturbation theory where instead of C*-algebras general topological $^*$-algebras 
are better suited.
Or one may use von Neumann algebras, in case particular states are
selected. On the other side, one might consider for $\Loc$ bundles over spacetimes,
or one might (in conformally invariant theories) admit conformal embeddings as morphisms.  
In case one is interested in spacetimes which are not globally hyperbolic one could look at the 
globally hyperbolic subregions (where care has to be payed to the causal convexity 
condition (i) above). 
 
Now we define the concept of locally covariant quantum field theory.
\begin{Def}${}$\\
$(i)$\quad
A {\bf locally covariant quantum field theory} is a
covariant functor $\euA$ from $\Loc$ to $\Obs$
and (writing $\alpha_{\psi}$ for $\euA(\psi)$)
with the covariance properties
$$ \alpha_{\psi'} \circ \alpha_{\psi} = \alpha_{\psi' \circ \psi}\,,
\quad \alpha_{{\rm id}_M} = {\rm id}_{\euA(M)}\,,$$
for all morphisms 
$\psi \in \Hom_{\Loc}(M_1,M_2)$, all morphisms $\psi' \in
$ \linebreak
 $\Hom_{\Loc}(M_2,M_3)$ and all $M \in
 \obj(\Loc)$.
\\[6pt]
$(ii)$\quad A locally covariant quantum field
theory described by a covariant functor $\euA$ is called {\bf causal}
if the following holds: Whenever there 
are morphisms $\psi_j \in \Hom_{\Loc}(M_j,M)$, $j
=1,2$, so that the sets $\psi_1(M_1)$ and $\psi_2(M_2)$ are causally
separated in $M$, then one has
$$ \left[
  \alpha_{\psi_1}(\euA(M_1)),\alpha_{\psi_2}(\euA(M_2))\right]
 = \{0\}\,, $$
where the element-wise commutation makes sense in $\euA(M)$.
\\[6pt]
$(iii)$\quad 
We say that a locally covariant quantum
field theory given by the functor $\euA$ obeys the {\bf time-slice axiom} if
$$ \alpha_{\psi}(\euA(M)) = \euA(M') $$
holds for all $\psi \in \Hom_{\Loc}(M,M')$ such that
$\psi(M)$ contains a Cauchy-surface for $M'$.
\end{Def}
Thus, a quantum field theory is an assignment of C$^*$-algebras to
(all) globally hyperbolic spacetimes so that the algebras are
identifiable when the spacetimes are isometric, in the indicated way.
This is a precise description of the {\bf generally covariant locality principle}.

\section{The traditional approach}

The traditional framework of algebraic quantum field theory, in the 
Araki-Haag-Kastler sense, on a fixed globally hyperbolic spacetime can be recovered 
from a locally covariant quantum field theory, 
i.e.\ from a covariant functor $\euA$ with
the properties listed above. 

Indeed, let $M$ be an object
in $\obj(\Loc)$. We denote by $\calK(M)$ the set of all open subsets
in $M$ which are relatively compact and contain with each pair of
points $x$ and $y$ also all $\gb$-causal curves in $M$ connecting $x$
and $y$ (cf.\ condition $(i)$ in the definition of $\Loc$). 
$O \in \calK(M)$, endowed with the metric of $M$
restricted to $O$ and with the induced orientation
and time-orientation is a member of $\obj(\Loc)$, and the
injection map $\iota_{M,O}: O \to M$, i.e.\ the
identical map restricted to $O$, is an element in
$\Hom_{\Loc}(O,M)$. With this notation it is easy to prove the
following assertion:
\begin{Thm} \label{localnet}
Let $\euA$ be a covariant functor with the above stated properties,
and define a map $\calK(M) \owns O \mapsto \calA(O) \subset
\euA(M)$ by setting
$$ \calA(O) := \alpha_{\iota_{M,O}}(\euA(O))\, .$$

Then the following statements hold:
\begin{itemize}
\item[(a)] The map fulfills isotony, i.e.
 $$O_1 \subset O_2 \Rightarrow
\calA(O_1) \subset \calA(O_2) \qquad \mbox{for all}\qquad  O_1,O_2 \in
\calK(M)\,.$$ 
\item[(b)] 
The group $G$ of isometric diffeomorphisms
$\kappa: M \to M$ (so that $\kappa_*\gb = \gb$) preserving orientation
and time-orientation, is represented 
by C$^*$-algebra automorphisms
${\alpha}_{\kappa} : \euA(M) \to \euA(M)$ 
such that
$$
{\alpha}_{\kappa}(\calA(O)) = \calA(\kappa(O))\,, \quad O \in
\calK(M)\,.
$$
\item[(c)] If the theory given by $\euA$ is additionally causal, then
  it holds that 
$$ [\calA(O_1),\calA(O_2)] = \{0\} $$
for all $O_1,O_2 \in \calK(M)$ with $O_1$ causally separated from
$O_2$.
\end{itemize}
\end{Thm} 
These properties are just the basic assumptions of the Araki-Haag-Kastler 
framework. 
\section{The achievements of the traditional approach}

In the Araki-Haag-Kastler approach in Minkowski spacetime $\mathbb M$ a 
great deal of results 
have been disclosed in the last forty years, some of those resulting in 
source of inspiration also to mathematics. 
Let us organize the description of the achievements in terms of a 
length-scale basis, from the small to the large. 
We assume in this section that the algebra $\euA(\mathbb M)$ is 
faithfully and irreducibly 
represented on a Hilbert space $\calH$, that the Poincar\'{e} transformations 
are unitarily implemented with positive energy, and that the subspace of 
Poincar\'{e} invariant vectors is one dimensional (uniqueness of the vacuum).
Moreover, algebras correponding to regions which are spacelike to a 
nonempty open region are assumed to be weakly closed 
(i.e. von Neumann algebras on $\calH$), and the condition of weak 
additivity is fulfilled, i.e. for all $O\in\calK(\mathbb M)$ 
the algebra generated from the algebras $\calA(O+x)$, $x\in\mathbb M$ is 
weakly dense  in $\euA(\mathbb M)$.  

\subsection{Ultraviolet structure and idealized localizations}
This part deals with the problem of inspecting the theory at very small 
scales. At the extreme we are interested in idealized localizations, eventually 
to points of spacetimes. But the observable algebras are 
trivial at any point $x\in \mathbb M$, namely 
$$
\bigcap_{O\ni x}\calA(O)=\bC {\mathbf 1}\ , \qquad O\in\calK(\mathbb M)\ .
$$

Hence pointlike localized observables are necessarily singular.
Actually, the Wightman formulation of quantum field theory 
is based on the use of distributions on spacetime with values in the algebra of 
observables (as a topological *-algebra). In spite of technical complications whose physical 
significance is unclear this formalism is well suited for a discussion of the connection 
with the euclidean theory which allows in fortunate cases a treatment by path integrals; it is 
more directly related to models and admits via the operator product expansion a study of the 
short distance behaviour. It is therefore an important question how the algebraic approach 
is related to the Wightman formalism. We refer to the literature for exploring the results on this
relation.

Whereas these results point to an essential equivalence of both formalisms, one needs in 
addition a criterion for the existence of sufficiently many Wightman fields associated to a 
given local net. Such a criterion can be given in terms of a compactness condition to be 
discussed in the next subsection. As a benefit one derives an operator product expansion 
which has to be assumed in the Wightman approach.

In the purely algebraic approach the ultraviolet 
structure has been investigated by Buchholz and Verch. Small scale
properties of theories are studied with the help of the so called
scaling algebras whose elements can be described as orbits of observables under
all possible renormalization group motions. There results
a classification of theories in the scaling limit which
can be synthetised into three broad classes; theories for which 
the scaling limit
is purely classical (commutative algebras), those for which the limit
is essentially unique (stable ultraviolet fixed point)
 and not classical and those for which this is 
not the case (unstable ultraviolet fixed point). This classification 
does not rely on perturbation expansions. It allows an intrinsic definition of confinement 
in terms of so-called 
ultraparticles, i.e. particles which are visible only in the scaling limit.

\subsection{Phase space analysis}
As far as finite distances are concerned there are two apparently competing
principles, those of nuclearity and modularity. The first one suggests that
locally, after a cutoff in energy, one has a situation similar to that of old quantum 
mechanics, namely, a finite
number of states in a finite volume of phase space. Aiming at a precise formulation  
Haag and Swieca introduced their notion of compactness, which Buchholz 
and Wichmann sharpened into that of nuclearity. The latter authors proposed that the 
set generated from the vacuum vector $\Omega$
$$
\{e^{-\beta H}A\Omega\ |\ A\in\calA(O)\ ,\ \|A\|<1\}\, ,
$$
$H$ denoting the generator of time translations (Hamiltonian), 
is nuclear for any $\beta>0$, roughly saying that it is contained in 
the image of the unit ball under a trace class operator. The
nuclear size $Z(\beta,O)$ of the set plays the r\^ole of the partition 
function of the model and has to satisfy certain bounds in 
the parameter $\beta$. The consequence of this constraint is the
existence of product states, namely those normal states for which
observables localized in two given space-like separated regions are uncorrelated.
A further consequence is the existence of thermal equilibrium states (KMS states) 
for all $\beta>0$. 

The second principle concerns the fact that 
even locally,
quantum field theory 
has infinitely many degrees of freedom. This
becomes visible in the Reeh-Schlieder Theorem which states that 
every vector $\Phi$ which is in the range of $e^{-\beta H}$ for
some $\beta>0$ (in particular the vacuum $\Omega$) 
is cyclic and separating for 
the algebras $\calA(O)$, $O\in\calK(\mathbb M)$, i.e. $\calA(O)\Phi$ is dense in 
$\calH$ ($\Phi$ is cyclic) and $A\Phi=0$, $A\in\calA(O)$ 
implies $A=0$ ($\Phi$ is separating). The pair $(\calA(O),\Omega)$
is then a von Neumann algebra in the so called \emph{standard form}. 
On such a pair the Tomita-Takesaki theory can be applied, namely 
the densely defined
operator
$$
SA\Omega=A^*\Omega\, ,\, A\in\calA(O)\, ,
$$
is closable, and the polar decomposition of its closure 
$\bar{S}=J\Delta^{1/2}$ delivers an antiunitary involution
$J$ (the modular conjugation) and a positive selfadjoint operator 
$\Delta$ (the modular 
operator) associated to the standard pair $(\calA(O),\Omega)$. These operators have the properties
$$
J\calA(O)J=\calA(O)'
$$
where the prime denotes the commutant, and
$$
\Delta^{it}\calA(O)\Delta^{-it}=\calA(O), \ t\in\mathbb R.
$$ 

The importance of this structure is based on the fact disclosed by 
Bisognano and Wichmann using Poincar\'e covariant 
Wightman fields and local 
algebras generated by them, that for specific regions in Minkowski spacetime
the modular operators have a geometrical meaning. Indeed, 
these authors showed for the pair $(\calA(W),\Omega)$ where $W$ denotes the wedge region
$W=\{x\in {\mathbb M}\ |\ |x^0|<x^1\}$
that the associated modular unitary $\Delta^{it}$ is 
the Lorentz boost with velocity $\tanh(2\pi t)$ in the direction $1$ and that
the modular conjugation $J$ is the $CP_1T$ symmetry operator with parity 
$P_1$ the
reflection w.r.t. the $x^1=0$ plane. 
Later, Borchers discovered that already on the purely algebraic level a corresponding structure exists.
He proved that given any standard pair 
$(\calA,\Phi)$ and a one-parameter group of unitaries 
$\tau\to U(\tau)$ acting
on the Hilbert space $\calH$ with a positive generator and such that
$\Phi$ is invariant and $U(\tau)\calA U(\tau)^*\subset\calA$, $\tau>0$, 
then the associated modular operators $\Delta$ and $J$ 
fulfil the commutation relations
\begin{eqnarray*}
\Delta^{it}U(\tau)\Delta^{-it}& = & U(e^{-2\pi t}\tau)\\
JU(\tau)J & = & U(-\tau)
\end{eqnarray*}
which are just the commutation relations 
between boosts and light-like translations.

Surprisingly, there is a direct connection between the two concepts of 
nuclearity and modularity. Indeed, in the nuclearity condition 
it is possible to replace the Hamiltonian operator by a 
specific function of the modular operator
associated to a slightly larger region. 
Furthermore under mild 
conditions nuclearity and modularity together determine the structure of 
local algebras completely; they are isomorphic to the unique hyperfinite type 
$\mathrm{III}_1$ von Neumann algebra.  

\subsection{Sectors, symmetries, statistics and particles}
Large scales are appropriate 
for discussing global issues like superselection sectors, statistics 
and symmetries as far as large spacelike distances are concerned and 
scattering theory, with the resulting notions of particles and 
infraparticles, as far as large timelike distances are concerned.

In purely massive theories where the vacuum sector has a mass gap and where the mass shell 
of the particles are isolated, a very satisfactory description of the multi-particle
structure at large times can be given. Using the concept of almost local particle generators, 
$$
\Psi=A(t)\Omega
$$
where $\Psi$ is a single particle state (i.e. an eigenstate of the mass operator), 
$A(t)$ is a family of almost local operators essentially localized in the kinematical 
region accessible from
a given point by a motion with the velocities contained in the spectrum of $\Psi$,  
one obtains the multi-particle states as limits of products $A_1(t)\cdots A_n(t)\Omega$ for 
disjoint velocity supports. The corresponding closed subspaces are invariant under 
Poincar\'{e} transformations and are unitarily equivalent to the Fock spaces of
non-interacting particles. 

For massless particles, no almost local particle generators can be expected to exist. 
In even dimensions, however, one can exploit Huygens principle to construct asymptotic 
particle generators which are in the commutant of the algebra of the forward or backward 
lightcone, respectively. Again, their products can be determined and deliver multiparticle states. 

Much less well understood is the case of massive particles in a theory which possesses also 
massless particles. Here, in general, the corresponding states are not eigenstates of the 
mass operator. Since QED as well as the standard model of elementary particles have this problem the 
correct treatment of scattering in these models is still under discussion. One attempt to a correct treatment
is based on the concept of so-called particle weights, i.e. unbounded positive functionals on a suitable algebra.
This algebra is generated by positive almost local operators annihilating the vacuum and 
interpreted as counters.

The structure at large spacelike scales may be analyzed by the theory of 
superselection sectors. The best understood case is that of locally 
generated sectors which are the objects of the DHR theory. Starting from a 
distinguished representation $\pi_0$(vacuum representation) which is assumed to fulfil 
Haag duality,
$$
\pi_0(\calA(O))= \pi_0(\calA(O'))'
$$ 
for all double cones $O$, one may look at all representations which are 
equivalent to the vacuum representation if restricted to the observables 
localized in double cones in the spacelike 
complement of a given double cone. Such representations give rise to 
endomorphisms of the algebra of observables, and the product of 
endomorphisms can be interpreted as a product of sectors (``fusion''). 
In general, these representations violate Haag duality, but there is a 
subclass of so-called finite statistics sectors where the violation of Haag duality is small,
in the sense that the nontrivial inclusion
$$
\pi(\calA(O))\subset \pi(\calA(O'))' \ ,
$$
has a finite Jones index.
These sectors form (in at least 3 spacetime dimensions) a symmetric 
tensor category with 
some further properties which can be identified, in a generalization of the Tannaka-Krein Theorem, as the dual of a unique 
compact group. This group plays the r\^{o}le of a global gauge group. The 
symmetry of the category is expressed in terms of a representation of the symmetric 
group. One may then enlarge the algebra of observables and obtains an algebra of 
operators which transform covariantly under the global gauge group and satisfy 
Bose- or Fermi commutation relations for spacelike separation.

In 2 spacetime dimensions one obtains instead braided tensor categories. They have been 
classified under additional conditions (conformal symmetry, complete rationality, $c<1$). 
To some extent they can be interpreted as duals of generalized quantum groups.

Concerning the question, whether all representations describing elementary particles are, 
in the massive case, DHR representations, one can show that in the case of a representation 
with an isolated mass shell there is an associated vacuum representation which becomes 
equivalent to the particle representation after restriction to observables localized 
spacelike to a given infinitely extended spacelike cone. This property is weaker than the 
DHR condition but allows in 4 spacetime dimensions the same construction of a global gauge 
group and of covariant fields with Bose- and Fermi commutation relations, respectively, 
as the DHR condition. In 3 space dimensions, however, one finds a braided tensor category, 
which has similar properties as those known from topological field theories in 3 dimensions.

The sector structure in massless theories is, due to the infrared problem, not well 
understood. This is in particular true for QED.
    
\section{Fields as Natural Transformations}

In order to be able to interpret the theory in terms of measurements one has to be able to 
compare observables associated to different regions of spacetime, or, even different 
spacetimes. In the absence of nontrivial isometries such a comparison can be made in terms 
of locally covariant fields. By definition these are natural transformations from the 
functor of quantum field theory to another functor on the category of spacetimes $\Loc$. 

The standard case is the functor which associates to every spacetime $M$ its space $\calD(M)$ 
of smooth compactly supported test functions.  There the morphisms are the pushforwards
$\calD\psi\equiv\psi_{*}$.     

\begin{Def}${}$\\
\ 
 A {\bf locally covariant quantum field} $\Phi$ is a
natural transformation between the functors $\euD$ and $\euA$, i.e., for 
any object
$M$ in $\Loc$ there exists a morphism $\Phi_{M}:\calD(M)\to
\euA(M)$ 
such that  
for any pair of objects $M_1$ and $M_2$ and any morphism $\psi$ 
between them, the following diagram
\begin{equation*}
\begin{CD}
\calD(M_1) @>\Phi_{M_1}>> \calA(M_1)\\
@V{\psi_*}VV     @VV{\alpha_{\psi}}V\\
\calD(M_2)@>>\Phi_{M_2}> \calA(M_2)
\end{CD}
\end{equation*}
commutes. 
\end{Def}

The commutativity of the diagram means, explicitely, that 
\begin{equation*}
\alpha_\psi\circ\Phi_{M_1} = \Phi_{M_2}\circ \psi_*
\end{equation*}
which is the seeked requirement for the covariance of fields. 
It contains, in particular the standard covariance condition for spacetime 
isometries.

Fields in the sense above are not necessarily linear. Examples for 
fields which are also linear are the scalar massive free 
Klein Gordon field on all globally hyperbolic spacetimes and its 
locally covariant Wick polynomials. In particular, the energy momentum tensors can be
constructed as locally covariant fields and they provide a crucial tool
for discussing the back reaction problem for matter fields. 

An example for the more general 
notion of a field 
are the local S-matrices in the St\"uckelberg-Bogolubov-Epstein-Glaser sense. 
These are unitaries $S_{M}(\lambda)$ 
with $M\in\Loc$ and $\lambda\in\calD(M)$ which satisfy the conditions
$$
S_M(0)=1
$$
$$
S_{M}(\lambda+\mu+\nu)=S_{M}(\lambda+\mu)S_M(\mu)^{-1}S_M(\mu+\nu)
$$ 
for $\lambda,\mu,\nu\in\calD(M)$ such that the supports of $\lambda$ and 
$\nu$ can be separated by a Cauchy surface of $M$ with $\supp\lambda$ in the 
future of the surface.

The importance of these S-matrices relies on the fact that they can 
be used to define a new quantum field theory. The new theory is 
locally covariant if the original theory was and if the local 
S-matrices satisfy the condition of a locally covariant field above. A
perturbative construction of interacting quantum field theory on 
globally hyperbolic spacetimes was completed in this way by Hollands and 
Wald, based on previous work by Brunetti and Fredenhagen.

\end{document}